\documentclass[aps,prd,twocolumn,groupedaddress,letterpaper,nofootinbib]{revtex4}

\usepackage{fullpage}
\usepackage{graphicx}
\usepackage{amsmath}
\usepackage{color}
\usepackage{comment}
\usepackage[margin=0.72in]{geometry}

\begin{document}

\title{Transients in finite inflation}

\author{Andrew Scacco}
\affiliation{Department of Physics, UC Davis}
\author{Andreas Albrecht}
\affiliation{Department of Physics, UC Davis}

\date{\today}

\begin{abstract}

We test a model of inflation with a fast-rolling kinetic-dominated initial condition against data from Planck using Markov chain Monte Carlo parameter estimation. We test both an $m^2 \phi^2$ potential and the $R+R^2$ gravity model and perform a full numerical calculation of both the scalar and tensor primordial power spectra. We find a slight (though not significant) improvement in fit for this model over the standard eternal slow-roll case.

\end{abstract}

\maketitle


\section{Introduction}
One of the greatest sources of data for modern cosmology is the cosmic microwave background radiation (CMB). This has been measured to extreme precision and the concordance model of cosmology has achieved tremendous success in matching the data. However, there are still small anomalies, one of which is a deficit of power in the CMB at low multipoles. A cutoff in the primordial power spectrum from inflation at small $k$ translates into a reduction of power in the CMB at low multipoles. This low-$\ell$ anomaly consists mainly of a slight dip in the power spectrum at $\ell \sim 20-40$ and a slightly reduced quadrupole, though because of cosmic variance the significance of the anomaly is not high. In light of this low-$\ell$ anomaly, there has been a great deal of interest in theories that predict a cutoff in the power spectrum~\cite{Jing:1994jw,Sinha:2005mn}. One of these theories is a kinetic-dominated (or fast-roll) start to slow-roll inflation, studied in~\cite{Ade:2015oja,Ade:2013uln,Cicoli:2014bja,Contaldi:2003zv,Destri:2009hn,Handley:2014bqa,Lello:2013awa,Linde:2001ae,Ramirez:2012gt,Ramirez:2011kk}.


The standard slow-roll model of inflation works by positing a new field, $\phi$, the inflaton, with some potential $V(\phi)$. One of the most common choices of potential and the one we use here is $V = \frac{1}{2} m^2 \phi^2$. The field rolls down the inflaton potential, obeying the equation of motion 
\begin{align}
\ddot{\phi}+3H\dot{\phi}+\frac{dV(\phi)}{d\phi} = 0,
\end{align}
where $H=\frac{\dot{a}}{a}$ is the Hubble parameter and $a$ is the scale factor.
The second term in this equation can be viewed as a friction term, and so the field can approach a terminal velocity and the field is in slow-roll. If the field starts very high up on a suitably chosen potential, it will rapidly approach slow-roll, and then will have a long period of slow-roll inflation until the field exits the slow-roll regime as it nears the minimum of the potential. 

It is possible that the effective field theory description of inflation breaks down near the Planck scale. In some models this manifests itself as a cutoff on the minimum value of the wavenumber $k$ in the primordial power spectrum~\cite{Jing:1994jw}. This cutoff in the primordial power spectrum also appears in theories with a finite number of total e-foldings of inflation. Our work is partially motivated by holographic ideas that suggest an upper bound to the total number of e-foldings of inflation~\cite{Phillips:2014yma,Albrecht:2014eaa,Albrecht:2011yg,Albrecht:2009vr,Banks:2003pt}, and in Sec.~\ref{sec:MCMC} we discuss how our model relates to those bounds.

In this work, we start the motion of the inflaton field on the potential in a period of fast-roll before slow-roll inflation, and vary the initial conditions of the field. We solve for the predicted CMB power spectrum using a full numerical approach, and determine which parameters are the best fit to the Planck data by varying them with a Markov chain Monte Carlo (MCMC) sampler. 

The Planck collaboration has just published its latest data and results~\cite{Ade:2015oja}, and has included an analytic model~\cite{Contaldi:2003zv} intended to be an approximation to the numerical model we analyze here. Our model is very similar, but more complete by not making as many approximations, and performing the full calculations for both the scalar and tensor primordial power spectra. Our more accurate model better tests the underlying theory and a comparison of our model to theirs\footnote{The authors of Ref.~\cite{Contaldi:2003zv} also perform an exact numerical calculation of the scalar primordial power spectrum to fit data from WMAP using a grid method but this exact numerical solution is not used by the Planck team.} can be found in Sec.~\ref{sec:MCMC}. Unfortunately, the improvements introduced in our model do not yield a significant improvement in the fit to the Planck data.

\section{Background}
We first consider a model of inflation with a potential $V=\frac{1}{2} m^2 \phi^2$. We numerically calculate the primordial power spectrum using the Mukhanov-Sasaki equation~\cite{Mukhanov:1985rz,Sasaki:1986hm} with a Bunch-Davies vacuum~\cite{Bunch:1978yq} at early times and kinetic-dominated initial conditions to start inflation.

Evolution of perturbations and the primordial power spectrum are given by the Friedmann equations and the Mukhanov-Sasaki equation:
\begin{align}
H^2 + \frac{K}{a^2} = \frac{1}{3}\left(\frac{1}{2}\dot{\phi}^2+V(\phi)\right)
\end{align}
\begin{align}
\ddot{\phi}+3H\dot{\phi}+\frac{dV(\phi)}{d\phi} = 0\\
\xi_k'' + \left(k^2-\frac{z''}{z}\right)\xi_k = 0 \label{eq:MS} \\
y_k'' + \left(k^2-\frac{a''}{a}\right)y_k = 0 \label{eq:MST} \\
z = \frac{\phi'}{H} = a \frac{\dot{\phi}}{H},
\end{align}
where $K$ is the curvature which we set to zero to require a flat universe, $\xi_k$ and $y_k$ are the scalar and tensor mode functions respectively for a mode of wavenumber $k$, and a prime represents the derivative with respect to conformal time $\eta$, while a dot represents the derivative with respect to proper time $t$. Here we are working in units where $\hbar = c = 8 \pi G = 1$. 

For the quadratic inflation potential in a flat universe with a fast-roll start where $\dot{\phi}^2 \gg m^2 \phi^2$, we choose the initial conditions of the background for the fast-roll start to inflation as in~\cite{Contaldi:2003zv}. We set the scale factor $a = 1$ and the conformal time $\eta = 0$ at the start of the numerical integration in the fast-roll regime:
\begin{align}
a = \sqrt{1 + 2 h \eta} \label{eq:aapprox} \\
\frac{z''}{z} = \frac{a''}{a} = \frac{-h^2}{(1+2 h \eta)^2}, \label{eq:zapprox}
\end{align}
where $h$ is the conformal time Hubble parameter at $\eta = 0$.

We assume initial conditions for the scalar and tensor mode functions as done in~\cite{Contaldi:2003zv}:
\begin{align}
\xi_k = y_k = \sqrt{\frac{\pi}{8 h}} \sqrt{1+2 h \eta} H_0^{(2)}\left(k \eta + \frac{k}{2 h}\right), \label{eq:IC}
\end{align}
where $H_0^{(2)}$ denotes the Hankel function of the second kind with index zero. These initial conditions have been obtained by solving Eqns.~\eqref{eq:MS} and~\eqref{eq:MST} assuming Eq.~\eqref{eq:zapprox} and requiring that we must have consistency with the predictions of inflation without a kinetic stage in the limit that the kinetic stage is pushed infinitely far into the past.

To make the calculation easier for large k, we make the reparameterization
\begin{align}
\xi \to X e^{-i k \eta}\\
y \to Y e^{-i k \eta},
\end{align}
which results in a set of transformed Mukhanov Sasaki equations
\begin{align}
X''-2 i k X' - \frac{z''}{z} X = 0\\
Y''-2 i k Y' - \frac{a''}{a} Y = 0,
\end{align}
with initial conditions
\begin{align}
X_0 = \xi_0\\
\dot{X}_0 = \dot{\xi}_0 + \frac{i k}{a_0} \xi_0\\
Y_0 = y_0\\
\dot{Y}_0 = \dot{y}_0 + \frac{i k}{a_0} y_0.
\end{align}

We choose to start the numerical integration in the kinetic dominated regime when $\dot{\phi_0}$ is one hundred times the value of $m \phi_0$. For convenience, in several equations we use $r_{init} \equiv \frac{\dot{\phi_0}}{m \phi_0}$.\footnote{Equations~\eqref{eq:aapprox}~and~\eqref{eq:zapprox} are correct in the $r \to \infty$ limit. In practical terms, the error caused by using these equations for $r=100$ is negligible.} In order to set up the numerical integration, we take the limit $\eta \to 0$ and the initial conditions are
%
\begin{align}
a(0) = a_0 = 1\\
\phi(0) = \phi_0\\
\dot{\phi}(0) = \dot{\phi_0} = r_{init} m \phi_0\\
h = \sqrt{\frac{r_{init}^2+1}{6}} m \phi_0\\
X_0 = Y_0 = \sqrt{\frac{\pi}{8 h}} H_0^{(2)}\left(\frac{k}{2 h}\right)\\
\dot{X_0} = \dot{Y_0} = \sqrt{\frac{\pi h}{8}} f\left(\frac{k}{2 h}\right)\\
f(x) = (1 + 2 i x) H_0^{(2)}(x)-2 x H_1^{(2)}(x).
\end{align}

\begin{figure}[htbp]
\begin{center}
\includegraphics[width=8cm]{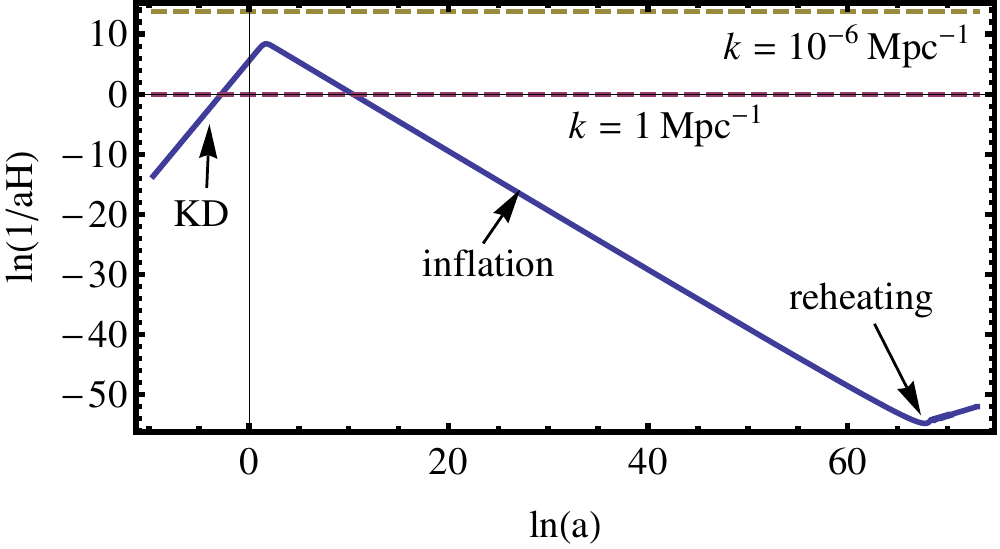}
\caption{Comoving Hubble length as a function of scale factor from the numerical code for a universe with an initial period of kinetic-dominated fast-roll (KD), followed by slow-roll inflation and reheating. In this plot, modes with wavenumber $k$ can be represented as horizontal lines, and a range of observable scales are shown. We have chosen this plot to correspond to the best fit parameters to the Planck data, $\phi_0 = 20.65$, and $m = 6 \times 10^{-6}$ corresponding to roughly 65 total e-foldings of inflation. \label{fig:avsaHinv}}
\end{center}
\end{figure}

We solve the Friedmann equations to get the background geometry (shown in Fig.~\ref{fig:avsaHinv}), and then use the results of this to find $z''/z$, from which we solve the Mukhanov-Sasaki equation where we use ordinary time, not conformal time, as the independent variable. Finally, we obtain the scalar and tensor power spectra for primordial perturbations:
\begin{align}
{\cal P}_{\cal R} = \frac{k^3}{2 \pi^2} \left| \frac{\xi_k}{z} \right|^2 \\
{\cal P}_t = \frac{k^3}{\pi^2} \left| \frac{2 y_k}{a} \right|^2.
\end{align}

\begin{figure}[htbp]
\begin{center}
\includegraphics[width=8cm]{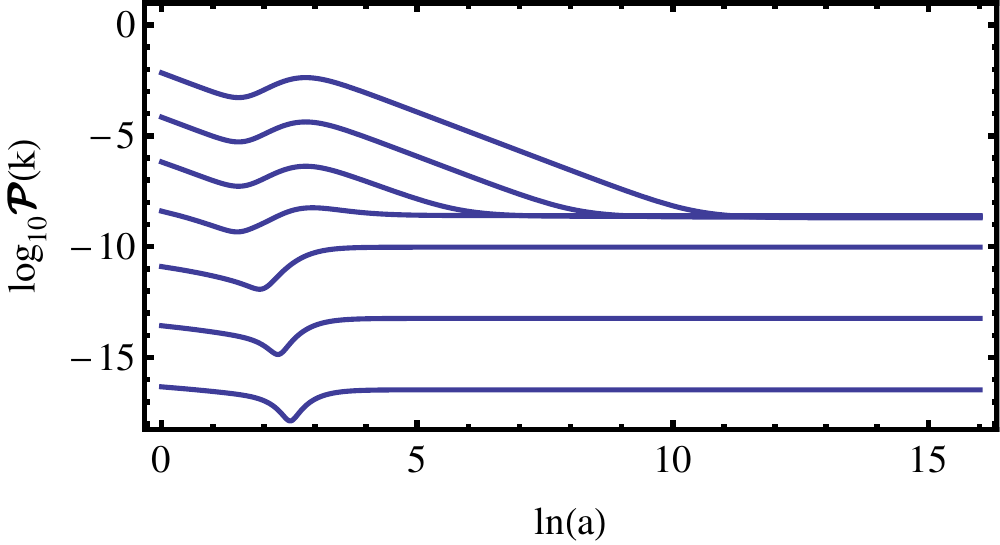}
\caption{The power in modes of ${\cal R}$ (shown for the best fit values of $m$ and $\phi_0$) as a function of scale factor for selected modes with $\log_{10}(k/{\rm Mpc}^{-1}) \in \{-6, -5, -4, -3, -2, -1, 0\}$ from lower to upper. The features near $\ln(a) \approx 2-3$ originate from the transition between the kinetic stage and slow-roll. We see from this plot that power is suppressed in modes with $\log_{10}(k/{\rm Mpc}^{-1}) \lesssim -3.5$. We show the case of standard slow-roll inflation in Fig.~\ref{fig:modeplotsr} for comparison. \label{fig:modeplot}}
\end{center}
\end{figure}

\begin{figure}[htbp]
\begin{center}
\includegraphics[width=8cm]{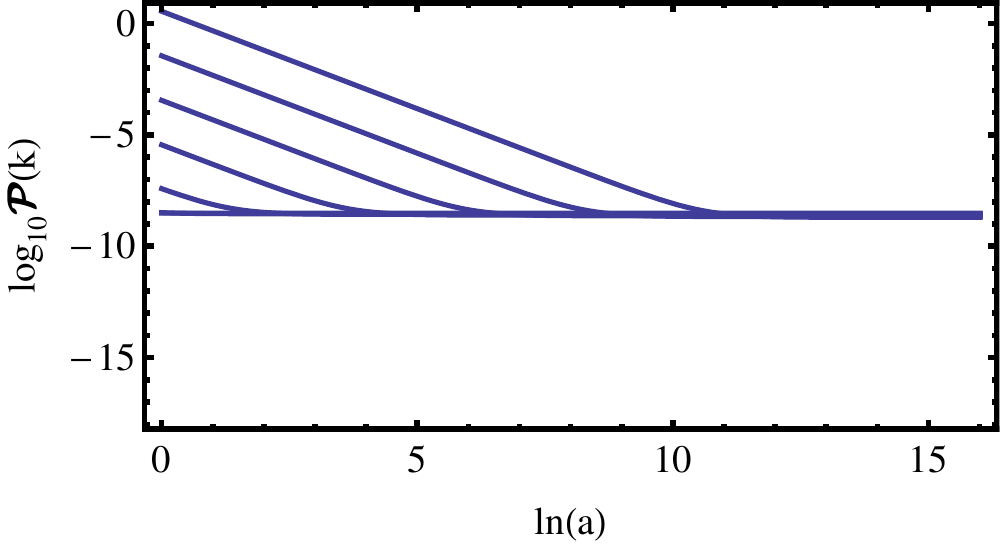}
\caption{The power in modes of ${\cal R}$ for standard slow-roll inflation as a function of scale factor for selected modes with $\log_{10}(k/{\rm Mpc}^{-1}) \in \{-6, -5, -4, -3, -2, -1, 0\}$ from lower to upper. The convergence of the curves to approximately the same value shows the approximate scale-invariance of the power spectrum. \label{fig:modeplotsr}}
\end{center}
\end{figure}

The evolution of modes of k can be seen from Fig.~\ref{fig:modeplot} and for comparison the same plot for standard slow-roll inflation is shown in Fig.~\ref{fig:modeplotsr}. When $k^2 \gg z''/z$, $\left| \xi_k \right|$ does not evolve, but $z$ does, and is roughly proportional to $a$ in both the fast-roll and slow-roll cases (though not during the transition between them), causing the downward slope of the large $k$ modes on Fig.~\ref{fig:modeplot}. When $k^2 \ll z''/z$, ${\cal R} = \frac{\xi_k}{z}$ does not evolve, and we say that the modes have frozen out. The integration may be stopped when the curvature perturbation ${\cal R}$ stops evolving (see Fig.~\ref{fig:modeplot}), which in the standard slow-roll model is usually assumed to happen after $k/aH \ll 1$, which is a good approximation for modes larger than the cutoff. For modes $k$ smaller than the cutoff, freeze-out of modes can be well approximated by the condition that the slow-roll parameter obeys $\eta \ll 1$~\cite{Kinney:2005vj}.\footnote{The general condition for freeze-out is always $k^2 \ll z''/z$.} For these small $k$ modes in our model, we find that $\eta \ll 1$ when $t \gg 1/m$ and stop the integration accordingly.

\begin{figure}[htbp]
\begin{center}
\includegraphics[width=8cm]{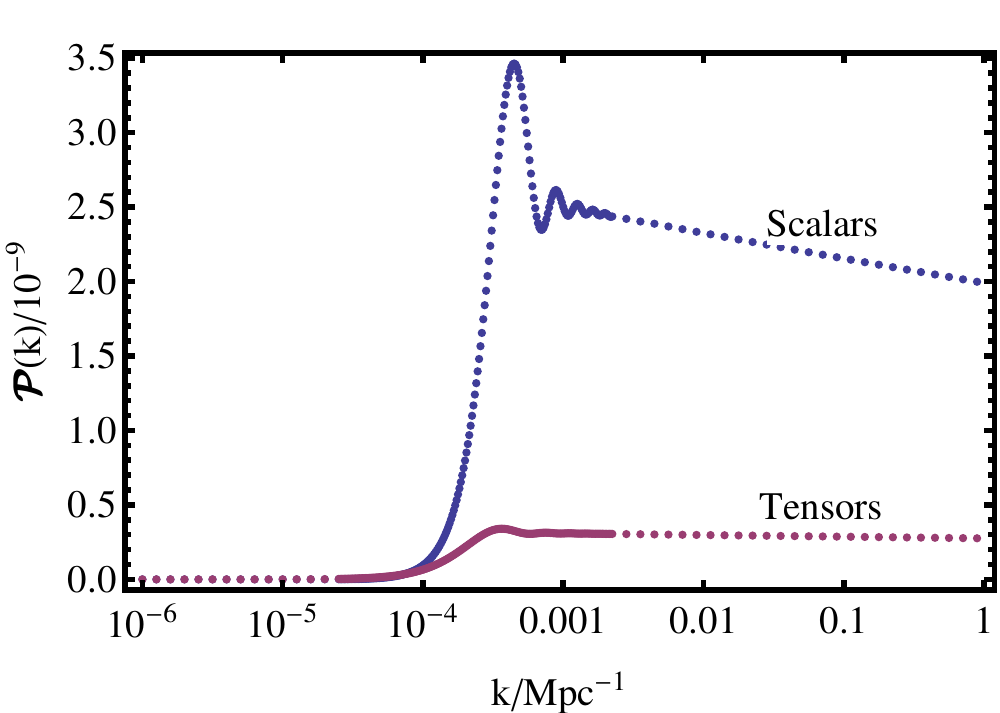}
\caption{The primordial power spectrum for scalars and tensors from the best fit numerical solution of the Mukhanov-Sasaki equation. Here $m_\phi = 6 \times 10^{-6}$ and $\phi_0 = 20.65$. \label{fig:pk}}
\end{center}
\end{figure}

The end of inflation is assumed to occur when $1/{aH}$ reaches a minimum. At this time, we assume instant reheating into a radiation dominated universe as shown in Fig.~\ref{fig:avsaHinv}. A longer period of reheating would impact the matching of the solution for the power spectrum from inflation onto the subsequent evolution of the universe. As such, it would result in a shift in the relationship between $\phi_0$ and the value of the cutoff in $k$. A longer period of reheating would correspond to fewer e-foldings of inflation, and hence a decrease in the effective value of $n_s$ for the part of the power spectrum with $k$ greater than the cutoff.\footnote{Since the constraints on $\phi_0$ are found to be weak in Sec.~\ref{sec:MCMC}, a full analysis of the impact of uncertainties in reheating was not performed.} An example of the power spectra for both scalars and tensors is shown in Fig.~\ref{fig:pk}.

%

\section{Numerical Implementation}
We solve the full Mukhanov-Sasaki equation starting from initial conditions, and solve for exact primordial power spectra for both scalars and tensors, and use this as numeric input for the Boltzmann code CLASS~\cite{Lesgourgues:2011re,Blas:2011rf} to calculate the $C_\ell$'s. We perform a full MCMC calculation, solving for the exact numerical solution at every choice of parameters. We find that the speed of the program is roughly halved by adding the numerical computation of scalar and tensor primordial power spectra.

We split the equations for the perturbations $X$ and $Y$ into real and imaginary parts, and integrate these equations together with the Friedmann equations for the evolution of the background simultaneously. For this purpose, we use a C implementation of a numerical differential equation solver that automatically switches between methods suitable for stiff and non-stiff equations (LSODA)~\cite{hind83} and run this solver from the command line in the external ${\cal P}(k)$ module~\cite{extpk} of CLASS. 

\begin{figure}[htbp]
\begin{center}
\includegraphics[width = 8cm]{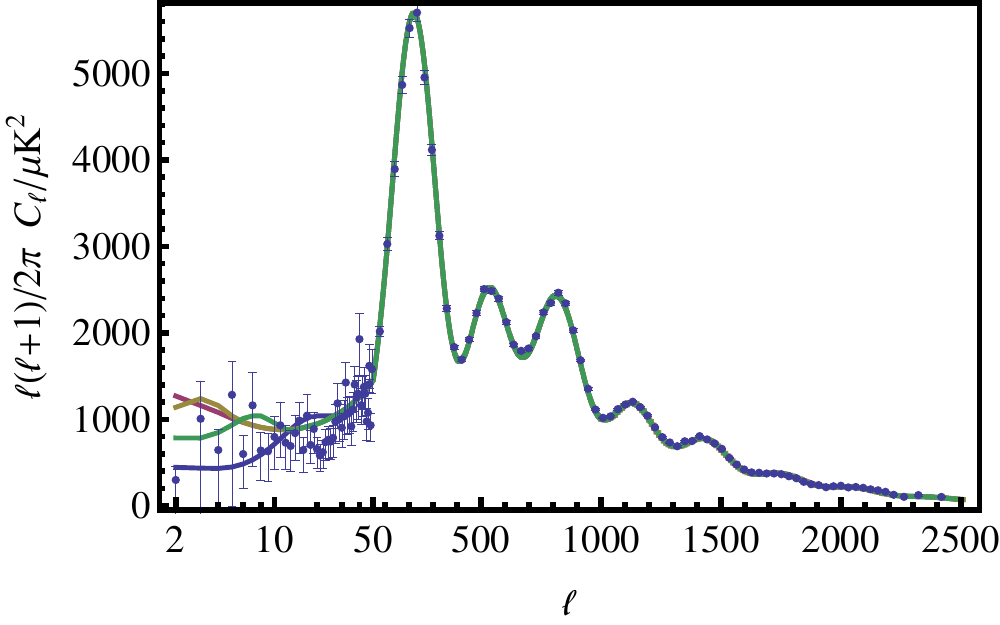}
\caption{A comparison with data of different predicted CMB power spectra with varying cutoffs (calculated using the CLASS Boltzmann code). The plots are all generated with $m_\phi = 6 \times 10^{-6}$ and where $\phi_0 \in \{20.5, 20.6, 20.7, 20.9\}$ (from the lowest curve to the highest on the left hand side). \label{fig:Clcompare}}
\end{center}
\end{figure}

\begin{figure}[htbp]
\begin{center}
\includegraphics[width = 8cm]{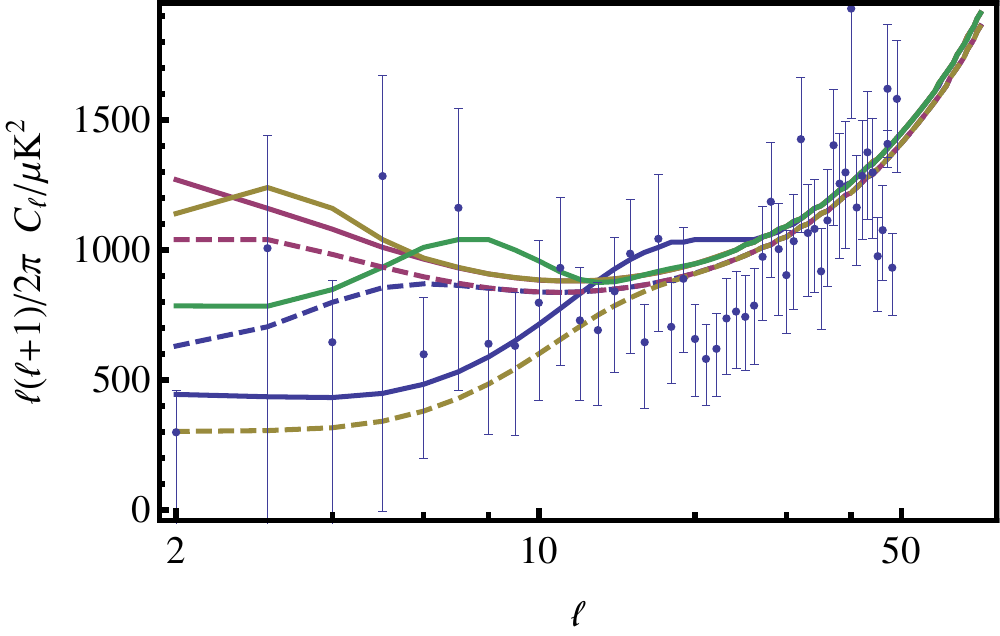}
\caption{A comparison with data of different predicted CMB power spectra with varying cutoffs (calculated using the CLASS Boltzmann code). The plots of the numerical primordial power spectra (solid) are all generated with $m_\phi = 6 \times 10^{-6}$ and where $\phi_0 \in \{20.5, 20.6, 20.7, 20.9\}$ (from the lowest curve to the highest on the left hand side). The dashed curves show the ansatz of Ref.~\cite{Contaldi:2003zv}, with $\ln(k_c/{\rm Mpc}^{-1}) \in \{-7, -8, -9\}$. Though all curves are capable of suppressing the lower multipoles, they are unable to reproduce the deficit of power seen in the data at $\ell \approx 20-40$ while preserving the good fit to the rest of the power spectrum. \label{fig:Clcomparelowl}}
\end{center}
\end{figure}

A comparison of our model to the data for the CMB power spectrum is shown in Figs.~\ref{fig:Clcompare} and~\ref{fig:Clcomparelowl}.

\section{MCMC} \label{sec:MCMC}
\begin{figure*}[htbp]
\begin{center}
\includegraphics[width = 17cm]{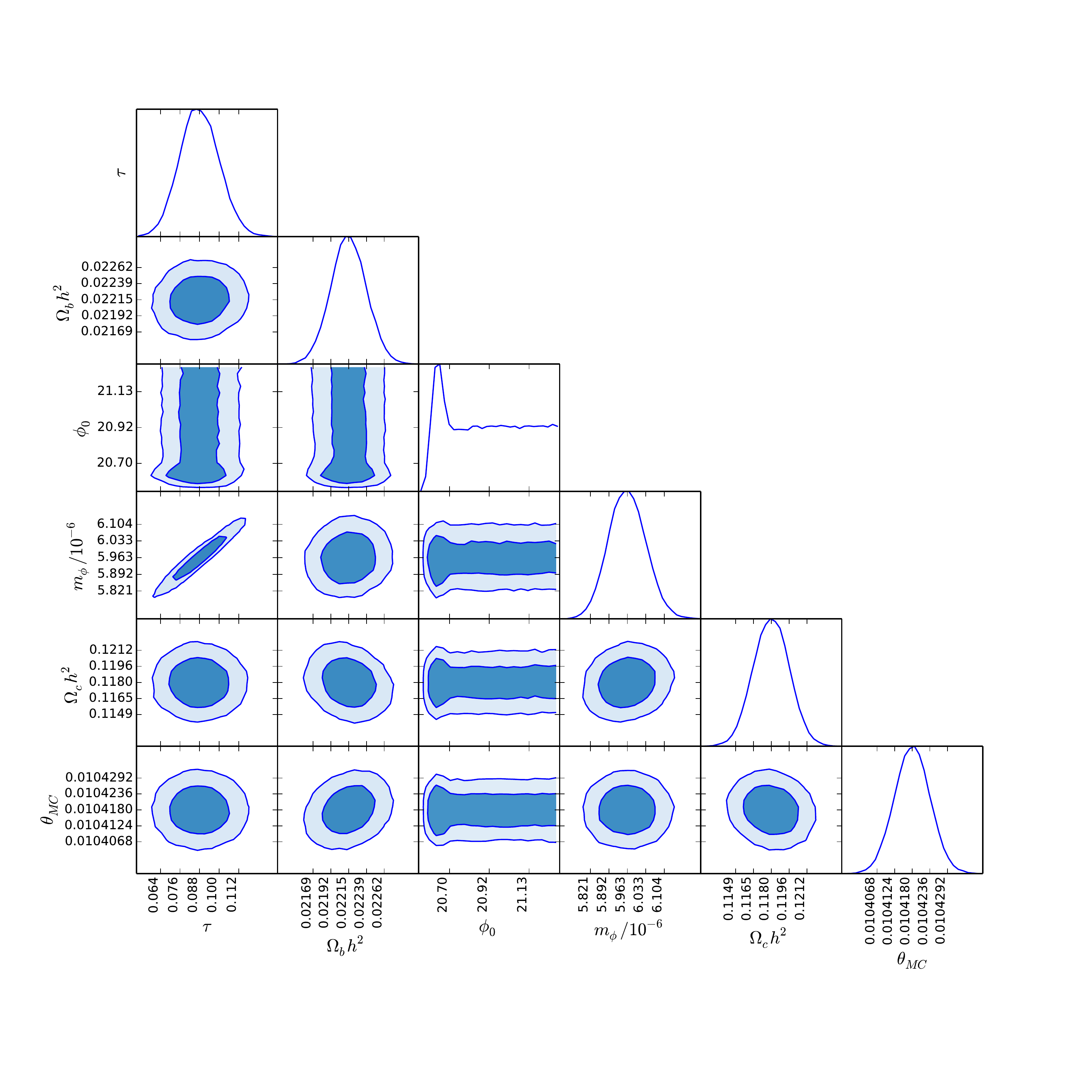}
\caption{The MCMC likelihood distributions of parameters. The mean values of parameters are displayed along with one and two sigma confidence intervals. The parameter $\phi_0$ is not sufficiently constrained. Small values of $\phi_0$ are ruled out as they would give too much power suppression, and large values would be observationally indistinguishable from standard slow-roll inflation (shown by the flat part of likelihood for $\phi_0$).  One can see in the 1-D $\phi_0$ plot that the likelihood for the best fit (at $\phi_0=20.65$) is only a factor of two larger than for the standard slow-roll inflation. \label{fig:likegrid}}
\end{center}
\end{figure*}

\begin{figure}[htbp]
\begin{center}
\includegraphics[width=8cm]{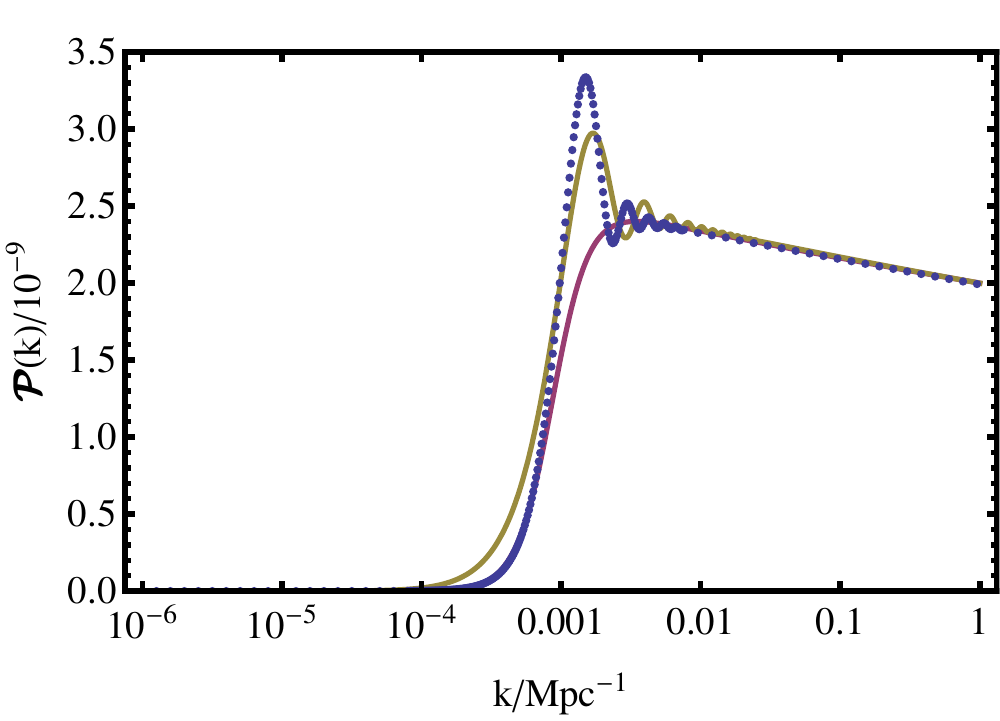}
\caption{The primordial power spectrum for scalars from the numerical solution of the Mukhanov-Sasaki equation (dotted), and compared to the ansatz of Contaldi {\it et al.} (lower solid) and the analytic model again from Contaldi {\it et al.} used in the Planck paper (upper solid). The dotted numerical curve has $m_\phi = 6 \times 10^{-6}$ and $\phi_0 = 20.5$.
\label{fig:CFvspkT}}
\end{center}
\end{figure}

\begin{figure}[htbp]
\begin{center}
\includegraphics[width=8cm]{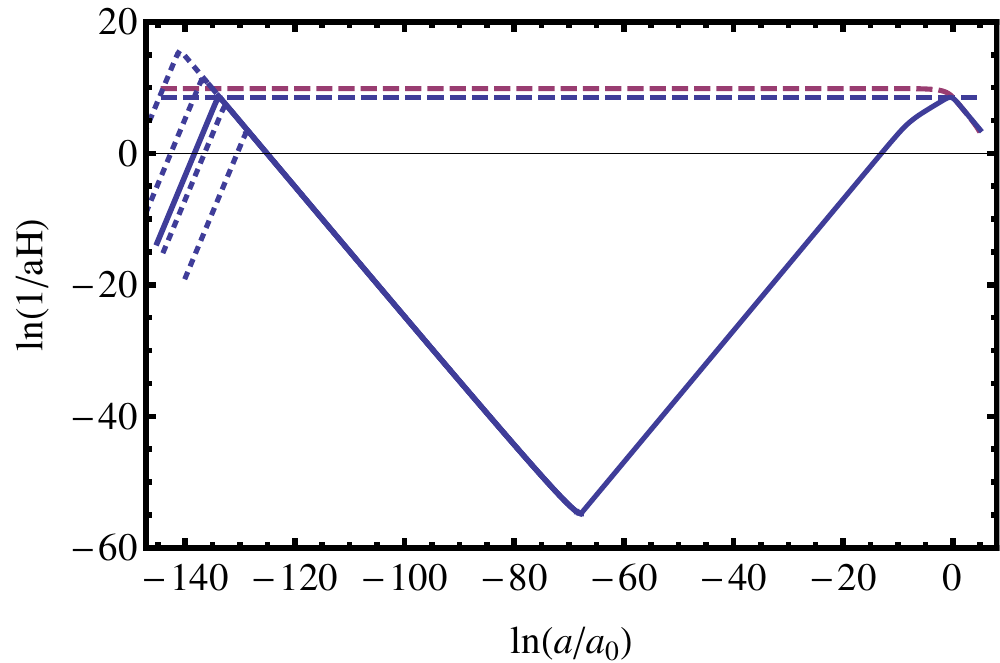}
\caption{Comoving Hubble length as a function of scale factor from the numerical code evolved from the fast-roll start to the end of inflation joined with the analytic solution for instantaneous reheating into radiation domination followed by matter domination and cosmological constant domination (solid). Increasing the initial value of the scalar field causes slow-roll inflation to begin earlier as shown by the dotted lines where $\phi_0 \in \{20.0, 20.5, 21.0, 21.5\}$, and $m = 6 \times 10^{-6}$. Changing the value of $m$ in a manner consistent with the observed perturbation amplitude results in subtle differences that are too small to be seen on the scale of this plot. Modes with wavenumber $k$ can be represented as horizontal lines, and the holographic bounds of Banks-Fischler (lower) and from de~Sitter Equilibrium (upper) are shown as dashed lines. The solid line corresponds to the best fit value of $\phi_0 \approx 20.65$ from the MCMC, and this value almost exactly saturates the Banks-Fischler bound. \label{fig:fullavsaHinv}}
\end{center}
\end{figure}

We run the MCMC using the CosmoSLik sampler~\cite{cosmoslik}, and vary over the values of $m$ and $\phi_0$, in addition to the standard $\Lambda$CDM parameters using the likelihoods for the temperature and low-$\ell$ polarization CMB power spectra from the Planck 2013 data. The results of the MCMC are shown in Fig.~\ref{fig:likegrid}. Because large values for $\phi_0$ are observationally indistinguishable from standard slow-roll inflation and in order for the chain to converge, we impose an upper bound on the value of $\phi_0$ shown in Fig.~\ref{fig:likegrid}. We find a slight preference for a kinetic start to inflation with the $m^2 \phi^2$ potential as opposed to always being in slow-roll. However, the likelihood is only a factor of two larger for the best fit cutoff model, and hence the improvement is not significant. The Planck collaboration~\cite{Ade:2015oja} showed that in a similar model~\cite{Contaldi:2003zv}, the expected improvement in log likelihood is not significant compared to the expected improvement from fitting an ensemble of cosmologies with perfect power laws as initial conditions and departures due solely to cosmic variance. The model used for the power spectrum in that case is similar to our exact numerical solution. Our numerical solution shows that there is a sharper suppression of power at small $k$, and more narrow oscillations of greater amplitude in the cutoff region as compared to the model of~\cite{Contaldi:2003zv} used in the Planck paper as shown in Fig.~\ref{fig:CFvspkT}. Also it appears that the kinetic start to inflation is unable to reproduce the shape of the observed dip in the CMB power spectrum in the neighborhood of $\ell \approx 20-40$. Interestingly, our best fit value of $\phi_0 \approx 20.65$ is very close to saturating the holographic bounds of Banks-Fischler~\cite{Banks:2003pt} and de~Sitter Equilibrium~\cite{Albrecht:2014eaa,Albrecht:2011yg,Albrecht:2009vr} shown in Fig.~\ref{fig:fullavsaHinv}. 

%
%
%


\section{$R+R^2$ inflation} \label{sec:r2}

We also adapt the method utilized for the $m^2 \phi^2$ potential to an $R+R^2$ gravity model~\cite{starobinsky_new_1980}. This model is the first inflationary model proposed, and is still an excellent fit to the Planck data.\footnote{Interestingly, by putting in a kinetic transient we deviate from the original spirit of~\cite{starobinsky_new_1980}. That paper offered a single ``correct'' solution for cosmology and explicitly rejected other solutions, such as the ones to which our transients belong, as uninteresting.} It has an action

\begin{align}
S = \int{d^4 x \sqrt{-g} \frac{M_{\rm pl}^2}{2}\left(R+\frac{R^2}{6 M^2}\right)}.
\end{align}

We work in the Einstein frame where the inflationary potential becomes

\begin{align}
V(\phi) = \Lambda^4 \left(e^{-\sqrt{\frac{2}{3}} \phi}-1\right)^2.
\end{align}

\begin{figure*}[htbp]
\begin{center}
\includegraphics[width = 17cm]{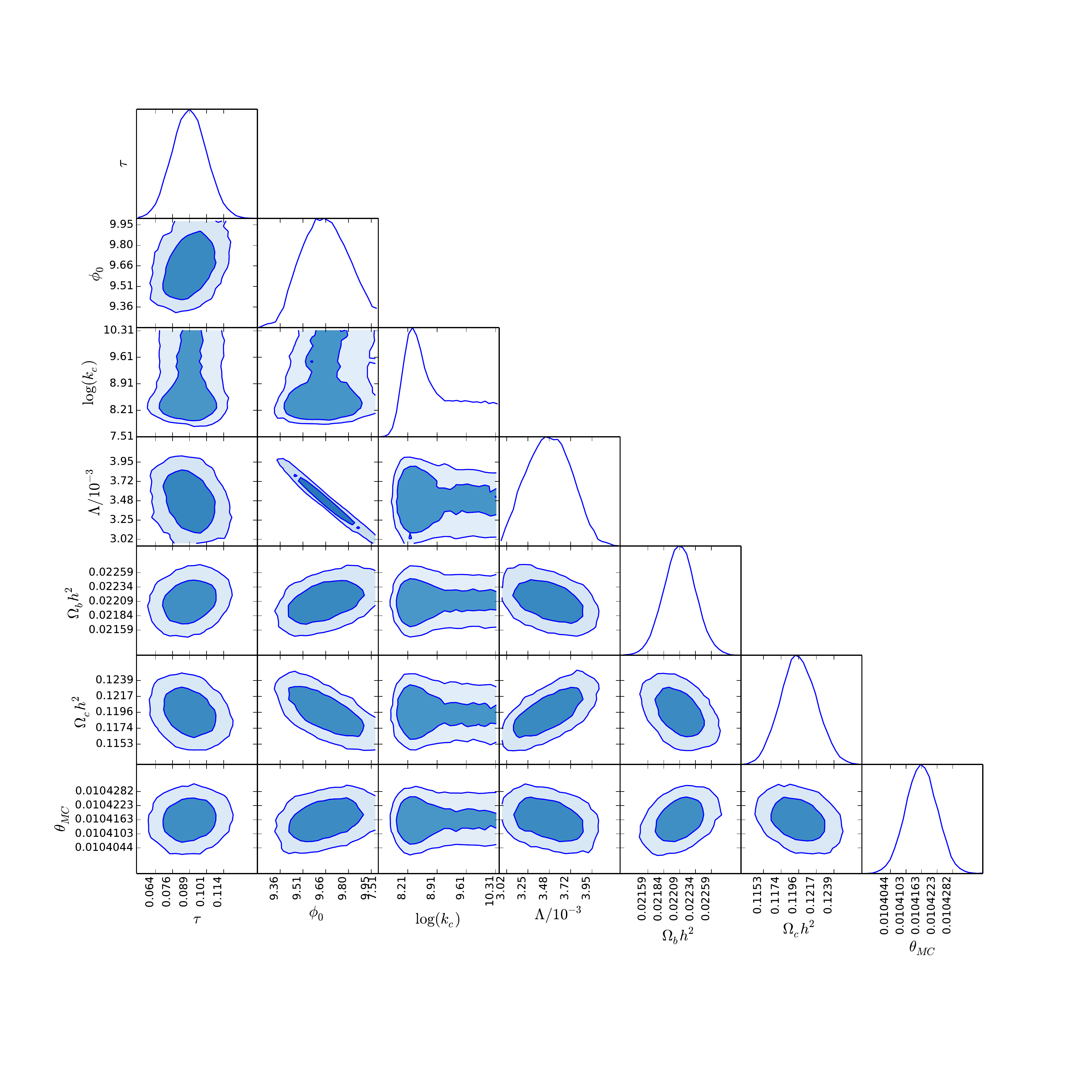}
\caption{The MCMC likelihood distributions of parameters for the $R+R^2$ inflation model. The mean values of parameters are displayed along with one and two sigma confidence intervals. The parameters $\Lambda$ and $\phi_0$ together determine both the amplitude of primordial perturbations and their effective spectral index. The parameter $k_c$ is not sufficiently constrained. Small values of $k_c$ are ruled out as they would give too much power suppression, and large values would be observationally indistinguishable from standard slow-roll inflation. As with $m^2 \phi^2$  inflation, the presence of a cutoff can at best increase the likelihood by a small factor relative to standard slow-roll inflation. \label{fig:likegridr2}}
\end{center}
\end{figure*}

This model is attractive due to its prediction of reduced tensor power. For this model, we also accommodate a variable length of reheating through varying an additional parameter $k_c$ corresponding to the minimum value of $aH$ which occurs during the transition from kinetic domination to slow-roll. We incorporate the same formalism used for the $m^2 \phi^2$ model and obtain parameter constraints on $\Lambda$, $k_c$, and $\phi_0$ as shown in Fig.~\ref{fig:likegridr2}.
Because of the reduced tensor power, this potential is a better fit to the Planck data than the $m^2 \phi^2$ potential; however we still do not obtain a significant improvement in the fit due to the presence of a fast-roll start to inflation. 


\section{Discussion}
Our work suggests that adding in a kinetic-dominated start to inflation does not significantly improve the fit to the CMB data from Planck. Despite this, there is a slight (though not significant) preference for a cutoff at approximately the value expected from a theory of finite inflation such as in~\cite{Banks:2015iya}.

Though our model has many similarities to the analytical model of Ref.~\cite{Contaldi:2003zv} used to fit the Planck data in~~\cite{Ade:2015oja}, we had hoped that by correcting for these differences in the MCMC we would improve the fit to the Planck data. However, it appears that the Planck data is not able to distinguish these differences, and thus our results do not significantly favor a cutoff in the primordial power spectrum.

\begin{acknowledgments}
We thank B. Follin, M. Millea, R. Holman, and A. Hernley for helpful discussions.  This work
was supported in part by DOE Grant DE-FG02-91ER40674.
\end{acknowledgments}

\bibliography{curve}
\bibliographystyle{apsrev}

\end{document}